# Continuous-wave Raman Lasing in Aluminum Nitride Microresonators

Xianwen Liu,[1] Changzheng Sun,[1,*] Bing Xiong,[1] Lai Wang,[1] Jian Wang,[1] Yanjun Han,[1] Zhibiao Hao,[1] Hongtao Li,[1] Yi Luo,[1] Jianchang Yan,[2] Tongbo Wei,[2] Yun Zhang,[2] and Junxi Wang[2]

[1]*Tsinghua National Laboratory for Information Science and Technology, State Key Lab on Integrated Optoelectronics, Department of Electronic Engineering, University, Beijing 100084, China*

[2]*Research and Development Center for Semiconductor Lighting, Institute of Semiconductors, Chinese Academy of Sciences, Beijing 100083, China*

We report the first investigation on continuous-wave Raman lasing in high-quality-factor aluminum nitride (AlN) microring resonators. Although wurtzite AlN is known to exhibit six Raman-active phonons, single-mode Raman lasing with low threshold and high slope efficiency is demonstrated. Selective excitation of $A_1^{TO}$ and $E_2^{high}$ phonons with Raman shifts of ~612 and 660 cm$^{-1}$ is observed by adjusting the polarization of the pump light. A theoretical analysis of Raman scattering efficiency within *c*-plane (0001) of AlN is carried out to help account for the observed lasing behavior. Bidirectional lasing is experimentally confirmed as a result of symmetric Raman gain in micro-scale waveguides. Furthermore, second-order Raman lasing with unparalleled output power of ~11.3 mW is obtained, which offers the capability to yield higher order Raman lasers for mid-infrared applications.

Stimulated Raman scattering (SRS) is an optical nonlinear frequency conversion process and has proved to be an efficient approach to extend the available spectral coverage of laser sources [1]. Since the Stokes light is frequency-shifted from the pump by an optical-phonon frequency, it features the potential to provide wavelength-agile tunable sources for sensing and spectroscopy applications. Compared with solid-state Raman lasers, microresonator-based configurations are particularly attractive, as they not only exhibit significantly reduced lasing threshold due to the field enhancement effect, but also enable device miniaturization [2–10]. Up to now, integrated Raman lasers with associated waveguides for pump and Stokes lights coupling have been experimentally demonstrated in both silicon [7, 8] and diamond [10]. Additionally, the capability of unidirectional lasing has been predicted in silicon as a result of the nonreciprocal Raman amplification in sub-micron waveguides [11–13].

Unlike silicon or diamond with a cubic crystal structure and a single optical phonon frequency at the center of Brillouin zone [14], wurtzite Aluminum nitride (AlN) belongs to space group $C_{6v}^4$ and provides six Raman-active phonons (namely $A_1^{TO}$, $A_1^{LO}$, $E_1^{TO}$, $E_1^{LO}$, $E_2^{low}$ and $E_2^{high}$ [15–18]) available for Raman lasing. Meanwhile, $A_1$ and $E_1$ are polar phonons exhibiting longitudinal-transverse optical (LO-TO) mode splitting as a consequence of the interaction between the lattice vibration and the long-range Coulomb field [15, 19]. The behavior of $E_2^{high}$ and $A_1^{LO}$ phonons recorded at *c*-plane surface of AlN has already been recognized as an efficient tool for monitoring the film crystalline quality, stress and free carrier concentration [18]. However, to our knowledge, the Raman lasing properties in AlN have not been explored up to now.

AlN features a wide direct bandgap (~6.2 eV at 300 K) as well as significant second- and third-order optical nonlinearities [20], making it attractive for applications in broadband nonlinear optics with negligible multi-photon absorption loss. Meanwhile, excellent thermal property and physical robustness allow AlN-based devices the high-power handling capacity. Currently, low waveguide propagation loss from visible to mid-infrared (MIR) region has been demonstrated with sputtered AlN [21, 22], together with nonlinear phenomena including second harmonic generation (SHG) [23], electro-optical modulation [24], and Kerr comb formation [25]. With improved crystalline quality and reduced grain-boundary size, strong Raman spectra with narrow-linewidth optic phonons are readily accessible when exploiting epitaxial AlN as the waveguide material. Furthermore, AlN grown on sapphire ($n_{sapphire}$ = ~1.75 at 1.55 μm) provides an ideal platform for guiding lights. Recently, AlN-on-sapphire microring with an intrinsic quality factor ($Q_{int}$) up to 3.2 million has been reported in our latest work [26]. In this Letter, we present the first demonstration of continuous-wave (CW) Raman lasing in AlN, and investigate the behaviors of Raman-active phonons involved in this nonlinear process.

In our experiment, a 1.2-μm-thick AlN film is grown on *c*-plane (0001) sapphire by metal organic chemical vapor deposition [27]. The *x*-ray diffraction (XRD) measurement reveals a full width at half maximum (FWHM) of ~47 arcsec along [0002] orientation, reflecting a high crystalline quality. To reduce the lasing threshold, high-*Q* microring resonators with integrated bus waveguides are fabricated following the process detailed elsewhere [26]. The microring has an outer radius of 80 μm and a width of ~3.5 μm, whereas the width of the bus waveguide is ~1.3 μm. The coupling gap between the microring and bus waveguide is 700 nm. The adopted microring dimensions ensure excellent spatial overlap between the pump and Stokes modes. To enhance the fiber-to-chip coupling efficiency and therefore promote the Raman laser output, the bus waveguide is tapered to a width of ~4 μm at both chip end facets.

To facilitate cleavage, the bus waveguide is chosen to run perpendicular to *m*-plane ($10\bar{1}0$) of AlN, which is the natural cleavage facet for hexagonal structures. For this purpose, the bus waveguide is defined perpendicular to the primary flat *a*-plane ($11\bar{2}0$) of the sapphire substrate (parallel to *m*-plane of the epitaxial AlN [28]). Figure 1 illustrates the waveguide end facet cleaved with a dicing machine after thinning the sapphire down to 150 μm and subsequent scribing with a pulsed UV laser [29]. It is noted that wrinkle-like features in the cleaved facet mentioned in Ref. [30] are greatly reduced here.

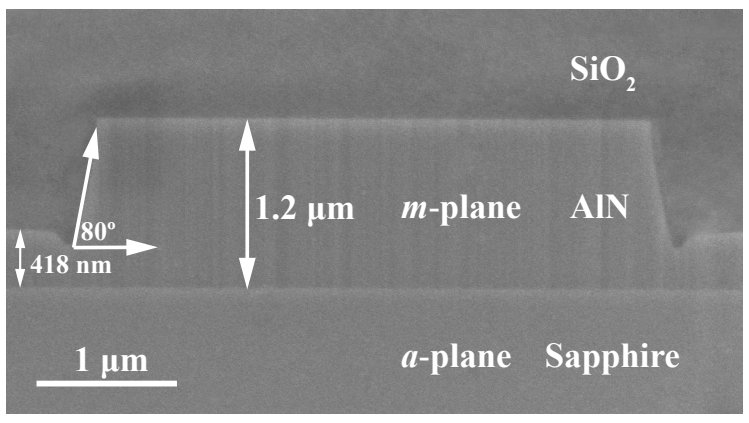

FIG. 1. Scanning electron microscopy (SEM) image of the cleaved waveguide facet (embedded with 3 μm silica). A pedestal structure is adopted with ~418 nm unetched AlN layer at the bottom, and the sidewall angle is ~80°. The waveguide is fabricated perpendicular to *a*-plane of sapphire, which is naturally aligned to *m*-plane of AlN.

The transmission spectrum shows that the microring supports two TM and two TE modes with a low off-resonance insertion loss of ~3.5 dB per facet. For both fundamental $TM_0$ and $TE_0$ modes, high on-resonance extinction ratios have been confirmed with free spectrum ranges (FSRs) of ~279 and 286 GHz, respectively. The measured loaded $Q$ factors are ~1.2 and 0.9 million for $TM_0$ and $TE_0$ modes, corresponding to intracavity power enhancement factors of ~530 and 342, respectively. The characterization of the device is detailed in Fig. S1 of Supplementary material.

To investigate the behavior of Raman-active phonons within the AlN waveguide, backscattering Raman spectroscopy measurement is carried out using a cleaved bus waveguide as the specimen (Fig. S2 in Supplementary material). Figure 2 shows the Raman peaks recorded with $x(z,\cdot)\bar{x}$ and $x(y,\cdot)\bar{x}$ scattering geometries, which are assigned to the corresponding optic phonons based on the reported Raman shifts [15–18]. Here, $A_1^{LO}$ and $E_1^{LO}$ phonons are prohibited for backscattering configurations at *m*-plane facet of AlN [19]. Additionally, unpolarized detection is employed here so as to permit direct comparison between the scattering efficiencies of excited phonons. To determine the polarization behavior of scattered lights, an additional polarizer is subsequently introduced. The linewidth of each phonon is extracted from a Lorentzian fit, and found to be consistent with the values reported for single-crystalline AlN [16]. The measurement results are summarized in TABLE I.

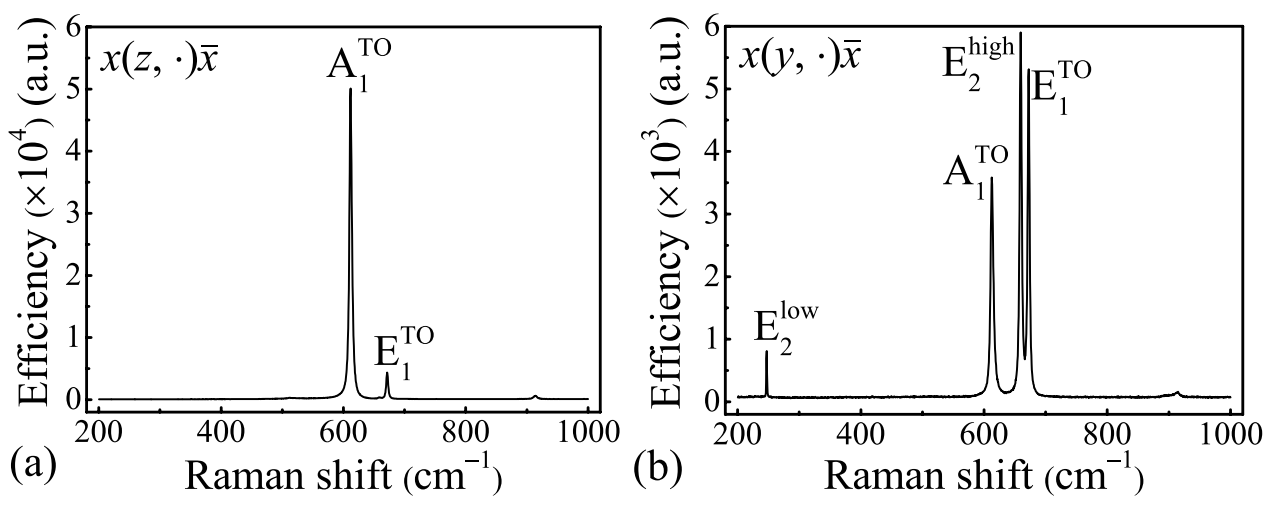

FIG. 2. (a) and (b) Recorded Raman spectra with $x(z,\cdot)\bar{x}$ and $x(y,\cdot)\bar{x}$ scattering geometries, respectively. The Cartesian coordinate is established with the $z$ and $x$ axes perpendicular to the *c*- and *m*-plane of AlN. The Porto notation is employed (the symbols outside the parentheses indicate the propagation directions of the incident and scattered lights, while those inside reveal their polarization directions). The dot (·) here means unpolarized detection.

TABLE I. Raman-active phonons at *m*-plane facet of AlN

| | Geometry | Linewidth (cm$^{-1}$) | Frequency (cm$^{-1}$) |
|---|---|---|---|
| $E_2^{low}$ | $x(yy)\bar{x}$ | 1.2 | 247.2 |
| $A_1^{TO}$ | $x(yy)\bar{x}$ | 5.1 | 612.6 |
| | $x(zz)\bar{x}$ | 4.6 | 611.6 |
| $E_2^{high}$ | $x(yy)\bar{x}$ | 3.8 | 659.4 |
| $E_1^{TO}$ | $x(zz)\bar{x}$ | 3.7 | 672.4 |
| | $x(zy)\bar{x}$ | 3.5 | 671.4 |

To enable Raman lasing, CW light from a tunable laser (Santec TSL-510, 1500−1630 nm) is boosted by a C-band erbium-doped fiber amplifier (EDFA) (1540−1570 nm), and gradually tuned into resonance from the blue-detuned side. SRS is promoted by the enhanced circulating power within the high-$Q$ microring, and Raman lasing subsequently occurs when the round-trip gain of the Stokes light overcomes its loss [31]. Furthermore, if the Stokes emission is close to the cavity resonance, highly efficient Raman laser output can be recorded accordingly. In our experiment, the Stokes light is monitored by an optical spectrum analyzer (OSA, Yokogawa AQ6375, 1200−2400 nm), whilst the pump is fine tuned to maximize the Stokes power.

Initially, the pump light from the EDFA is aligned to TM polarization (i.e., along *c*-axis of AlN), akin to $x(z,\cdot)\bar{x}$ geometry in Fig. 2(a). Therefore, $A_1^{TO}$ and $E_1^{TO}$ phonons are expected to be involved in the lasing process. However, only $A_1^{TO}$ mode with a Raman shift of ~612 cm$^{-1}$ is observed [Fig. 3(a)], revealing the single-mode lasing action. Meanwhile, a low-noise operation with a side-mode suppression ratio (SMSR) over 70 dB is confirmed. According to the Raman selection rules given in TABLE I, the Stokes light is inferred to be predominately TM-polarized. The threshold for first-order (1$^{st}$) Stokes is as low as ~8 mW [inset of Fig. 3(b)], as a result of significant intracavity power enhancement in the microring. The output 1$^{st}$ Stokes power (recorded by OSA hereinafter) exhibits a linear dependence on the pump and the unidirectional slope efficiency is ~3.6%.

Thanks to the low lasing threshold, wideband discrete tuning of 1$^{st}$ Stokes wavelength is feasible by employing a high-power tunable laser (Fig. S3 in Supplementary material), which allows extending the Raman lasing wavelength restricted by the EDFA bandwidth. Furthermore, there is the potential to realize continuously electro-optically tunable Raman sources by taking advantage of the intrinsic second-order (2$^{nd}$) optical nonlinearity of AlN [24], which is inaccessible in centrosymmetric silicon or diamond and can be valuable for practical applications. By further increasing the pump power, the intracavity 1$^{st}$ Stokes power will be high enough to act as a secondary pump source to enable 2nd Stokes lasing at ~1928.7 nm [Fig. 3(c)]. The threshold for 2nd Stokes lasing is as high as ~602 mW [inset of Fig. 3(d)]. Nevertheless, high unidirectional slope efficiency of ~5.4% is maintained. Moreover, an unparalleled 2nd Stokes output power of ~11.3 mW is achieved with ~794 mW pump, exceeding the value in silicon racetrack resonator [8]. The high-power 2nd Raman lasing in AlN suggests the possibility to realize higher order Raman lasers for mid-infrared applications [32].

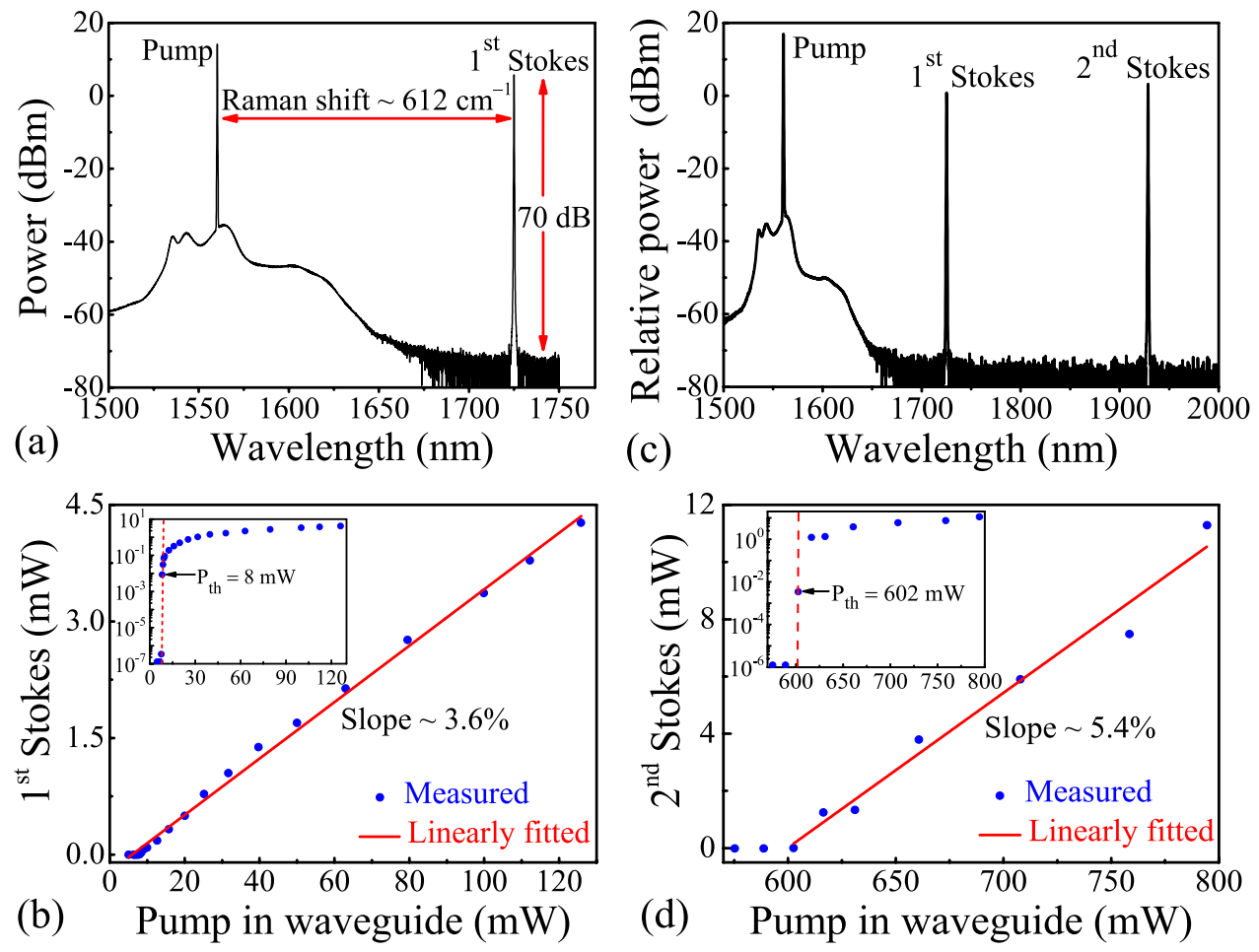

FIG. 3. Raman lasing in AlN with TM-polarized pump. (a) 1st Stokes lasing around 1725 nm with a Raman shift of ~612 cm$^{-1}$ from the pump (~1560.3 nm) with ~126 mW power. (b) The output 1st Stokes power exhibits linear dependence on the pump, and the unidirectional slope efficiency is ~3.6%. Inset: a log plot of the Stokes power with ~50 dB jump once the pump power is above threshold (~8 mW). (c) Observed 2nd Stokes lasing at ~1928.7 nm with an increased pump power of ~794 mW (7 dB attenuation is employed for the pump before the OSA). (d) Recorded output 2nd Stokes power versus the pump. The lasing threshold is ~602 mW, whereas the unidirectional slope efficiency is up to ~5.4%.

To investigate Raman lasing associated with other optic phonons, the pump is adjusted to TE polarization, analogous to $x(y,\cdot)\bar{x}$ geometry in Fig. 2(b). Hence, $E_2^{low}$, $A_1^{TO}$, $E_1^{TO}$ and $E_2^{high}$ phonons may contribute to the lasing process. Nevertheless, single-mode lasing with a Raman shift of ~660 cm$^{-1}$ is observed [Fig. 4(a)], indicating that only $E_2^{high}$ phonon is excited. Meanwhile, the selection rules in TABLE I require the Stokes emission to be quasi-TE polarized. Additionally, single FSR-spaced sidebands around 1st Stokes light are noticed, albeit with a notably low power. Since the Raman linewidth of $E_2^{high}$ phonon (~3.8 cm$^{-1}$ or ~114 GHz) is smaller than the microring FSR (~286 GHz), these sidebands actually fall out of the Raman gain bandwidth. Consequently, this phenomenon is inferred to be Raman induced hyperparametric oscillation through Kerr nonlinearity, as the microring has an anomalous dispersion and high intracavity power at 1st Stokes wavelength.

In contrast to the TM-polarized case, 2nd Stokes lasing around 1963.4 nm is recorded at a relatively low pump power (~126 mW). Actually, the lasing thresholds for 1st and 2nd Stokes lights are very close (~34 and 40 mW, respectively) [insets of Figs. 4(b) and 4(c)]. The unidirectional slope efficiency for 1st Stokes lasing is as high as ~15%, whereas that for 2nd Stokes light is much lower (~0.037%). Moreover, a high 1st Stokes output power of ~13.5 mW is obtained for ~126 mW pump. A further increase in pump (~158 mW) fails to enhance the 2nd Stokes power as expected. Instead, Raman-assisted four-wave mixing occurs with satellite sidebands spaced by 10 FSR around the pump, anti-Stokes, and 1st and 2nd Stokes lights [Fig. 4(d)], which is analogous to the prior observation in silica [2, 3]. Moreover, an evident increase in 1st anti-Stokes power is noticed, which may be ascribed to the coherent anti-Stokes Raman scattering (CARS) as a result of the decreased phase mismatch with increased pump [33]. This phenomenon can be of particular interest, as it offers the potential to realize on-chip deep UV Raman laser in AlN through amplification of anti-Stokes light, which is distinct from traditional sum-frequency approaches [34].

It has been previously predicted that unidirectional Raman lasing is accessible for sub-micron silicon waveguides via the nonreciprocal Raman amplification [11–13], resulting in Stokes power output at only one end facet. To investigate the directionality of AlN Raman laser, an additional 3-dB coupler is inserted between the EDFA and the microring chip, so as to allow the backward propagating light to be collected by the OSA. As illustrated in Figs. 5(a) and 5(b), strong back-ward Stokes lights are observed for both TM and TE pump, which confirms the symmetric Raman gain for both directions in micro-scale AlN waveguides [35]. Here, the absence of 2nd Stokes emission for TM pump is mainly attributed to its large threshold and the attenuation of pump power caused by the 3-dB coupler. It should be mentioned that there exists ∼11% reflection at the chip end facets, as shown in the Fabry-Pérot interference fringes of the amplifier spontaneous emission (ASE) noise. Hence, both the observed forward and backward Stokes lights contain 11% power components of their counterparts.

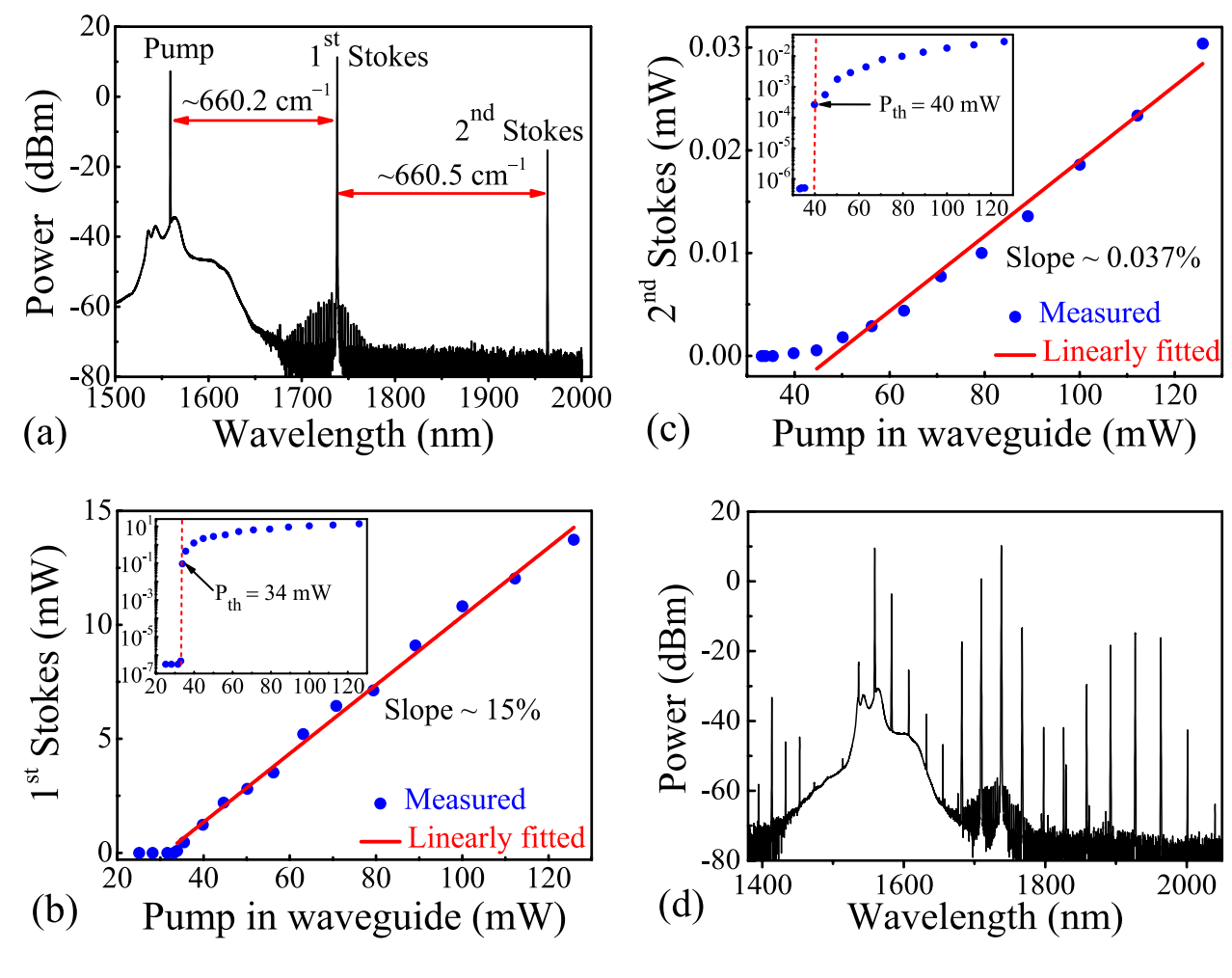

FIG. 4. Raman lasing in AlN with TE-polarized pump. (a) 1st and 2nd Stokes lights at ~1738 and 1963.4 nm with a Raman shift of ~660 cm$^{-1}$ for pump at ~1559.1 nm with ~126 mW power. (b) and (c) the output 1st and 2nd Stokes power versus the pump. Inset: a log plot of the Stokes power with the estimated thresholds of ~34 and 40 mW for 1st and 2nd Stoke emission, respectively. A high unidirectional output slope efficiency of ~15% is recorded for 1st Stoke lasing, whereas 2nd Stokes emission exhibits a much lower value (~0.037%). (d) Raman-assisted four-wave mixing is observed near the pump, Stokes and anti-Stokes lights when the pump is increased to ~158 mW.

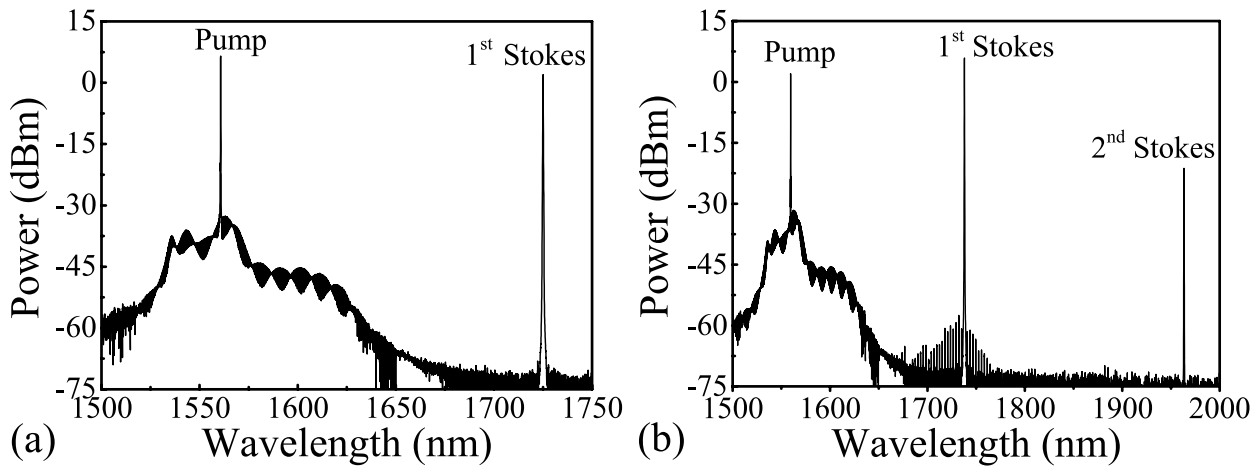

FIG. 5. Backward emission spectra for (a) TM- and (b) TE-polarized pump with ∼23 dBm power. The observed backward pump (with reduced power) is induced by the waveguide end facet reflection.

Although there are six Raman-active phonons in AlN, highly efficient single-mode Raman lasing is demonstrated for both TM- and TE- polarized pump, corresponding to the selective excitation of $A_1^{TO}$ and $E_2^{high}$ phonons, respectively. It is well known that the stimulated Raman gain coefficient $g_R$ is determined by the spontaneous scattering efficiency $S$ and Raman linewidth Γ [36]:

$$g_R = \frac{8\pi^2 c^2}{\hbar\omega_s^3}\frac{S}{n_s^2(N_0+1)\Gamma} \tag{1}$$

Here, $\omega_s$ and $n_s$ are the angular frequency and refractive index of Stokes lights, respectively. Meanwhile, $c$ is the light speed in vacuum, and $N_0$ is the Bose factor denoting the number density of Raman-active phonons. Hence, it is possible to relate the observed lasing behavior to the scattering efficiency $S$ of the involved phonons [15, 37, 38]:

$$S = S_0|\boldsymbol{e}_s \cdot \boldsymbol{R} \cdot \boldsymbol{e}_i|^2 \tag{2}$$

where $S_0$ is a constant of proportionality, $\boldsymbol{R}$ represents the Raman tensor, $\boldsymbol{e}_i$ and $\boldsymbol{e}_s$ describe the polarization vectors of incident and scattered lights, respectively.

According to the theoretical analysis detailed in section IV of Supplementary material, for TM or TE pump light propagating within the $c$-plane of AlN, the scattering efficiency for the optic phonons is independent of the waveguide orientation. Consequently, the Raman gain coefficient in the microring can be inferred from the measured spontaneous Raman spectra in Fig. 2. It turns out that maximum scattering efficiencies for $A_1^{TO}$ and $E_2^{high}$ phonons are attainable for TM and TE pump, respectively (Fig. S5 in Supplementary material). Meanwhile, taking into account the Raman linewidth in TABLE I, it is readily found from Eq. (1) that $A_1^{TO}$ and $E_2^{high}$ phonons also exhibit the highest $g_R$. Hence, the observed single-mode Raman lasing in our experiment can be interpreted as the consequence of mode competition in AlN microring. Additionally, the deduced scattering efficiency for $A_1^{TO}$ with TM pump case is ∼10.3 times than that of $E_2^{high}$ phonon, which helps account for its lower 1st Stokes lasing threshold.

In microresonator-based configurations, it should be mentioned that efficient Raman lasing also depends on aligning both pump and Stokes lights to their respective cavity resonances. In such a doubly resonant microcavity, deviation of Stokes wavelength from the resonance would critically affect the observed lasing threshold and slope efficiency. Thus, the low output slope efficiency of 2nd Stokes lasing with TE pump [Fig. 4(c)] is inferred to be induced by a severe deviation from the cavity resonance. This can be confirmed by its low lasing threshold, which implies an significantly reduced output coupling loss.

Based on the 1st Stokes lasing threshold formula [2, 3], the extracted Raman gain coefficient in AlN are ∼0.45 and 0.25 cm/GW for TM and TE pump, respectively (section V in Supplementary material), which is about ten times lower than that in diamond [10] and silicon [39]. However, highly efficient Raman lasing in AlN microring is experimentally achieved benefiting from the significant intracavity power enhancement as well as the excellent alignment between the Stokes wavelength and the cavity resonance. Meanwhile, the realization of microring-based AlN Raman lasers with a high efficiency is also a consequence of the waveguide-orientation independent scattering efficiency within $c$-plane of AlN for TM or TE pump, which is in contrast to that in silicon [7, 8, 35] and diamond [10] with racetrack resonators. Additionally, taking into account the measured linewidth of ∼4.6 and 3.8 $cm^{-1}$ as well as the deduced $S$ ratio of ∼10.3 for $A_1^{TO}$ and $E_2^{high}$ phonons, the ratio between $g_R$(TM) and $g_R$(TE) is derived to be ∼8.5 from Eq. (1), which is in reasonable agreement with our experimentally extracted one (∼1.8).

In conclusion, AlN-on-sapphire microring resonator has been demonstrated as a novel platform for highly efficient CW Raman lasing. Single-mode Raman laser with high SMSR is confirmed for both TM- and TE-polarized pump lights, corresponding to the selective excitation of $A_1^{TO}$ and $E_2^{high}$ phonons, respectively. Bidirectional Raman lasing is experimentally observed, attributed to the symmetric Raman gain in micro-scale AlN waveguides. The Raman scattering efficiency within $c$-plane of AlN is found to be independent of the waveguide orientation for both TM and TE pump, and should provide a generic design guide for on-chip AlN Raman lasers. Apart from SRS, other nonlinear phenomena occur at sufficiently high TE pump power, which hampers the Stokes power enhancement and will be the topic for further analysis. Additionally, AlN-based Raman laser features the potential to realize continuously electro-optically tunable Raman sources thanks to its intrinsic 2nd optical nonlinearity [24].

### ACKNOWLEDGEMENT

This work was supported in part by National Basic Research Program of China (2012CB315605, 2014CB340002), National Natural Science Foundation of China (61210014, 61321004, 61307024, 61574082, 51561165012), High Technology Research and Development Program of China (2015AA017101), Tsinghua University Initiative Scientific Research Program (20131089364, 20161080068, 20161080062), and Open Fund of State Key Laboratory on Integrated Optoelectronics (IOSKL2014KF09). We would like to thank Prof. Changxi Yang of Tsinghua University for the loan of optical spectrum analyzer (Yokogawa AQ6375).

---

* czsun@tsinghua.edu.cn

[1] H. M. Pask, Prog. Quant. Electron. **27**, 3 (2003).
[2] S. M. Spillane, T. J. Kippenberg, and K. J. Vahala, Nature **415**, 621 (2002).
[3] T. J. Kippenberg, S. M. Spillane, D. K. Armani, and K. J. Vahala, Opt. Lett. **29**, 1224 (2004).
[4] T. J. Kippenberg, S. M. Spillane, B. Min, and K. J. Vahala, IEEE J. Sel. Top. Quant. **10**, 1219 (2004).
[5] I. S. Grudinin and L. Maleki, Opt. Lett. **32**, 166 (2007).
[6] F. Vanier, M. Rochette, N. Godbout, and Y.-A. Peter, Opt. Lett. **38**, 4966 (2013).
[7] H. S. Rong, S. B. Xu, Y.-H. Kuo, V. Sih, O. Cohen, O. Raday, and M. Paniccia, Nat. Photonics **1**, 232 (2007).
[8] H. S. Rong, S. B. Xu, O. Cohen, O. Raday, M. Lee, V. Sih, and M. Paniccia, Nat. photonics **2**, 170 (2008).
[9] Y. Takahashi, Y. Inui, M. Chihara, T. Asano, R. Terawaki, and S. Noda, Nature **498**, 470 (2013).
[10] P. Latawiec, V. Venkataraman, M. J. Burek, B. J. Hausmann, I. Bulu, and M. Lončar, Optica **2**, 924 (2015).
[11] M. Krause, H. Renner, and E. Brinkmeyer, Appl. Phys. Lett. **95**, 261111 (2009).
[12] J. Müller, M. Krause, H. Renner, and E. Brinkmeyer, Opt. Express **18**, 19532 (2010).
[13] M. Krause, J. Appl. Phys. **111**, 093107 (2012).
[14] R. Loudon, Adv. Phys. **13**, 423 (1964).
[15] L. E. McNeil, M. Grimsditch, and R. H. French, J. Am. Ceram. Soc. **76**, 1132 (1993).
[16] M. Bickermann, B. M. Epelbaum, P. Heimann, Z. G. Herro, and A. Winnacker, Appl. Phys. Lett. **86**, 131904 (2005).
[17] T. Prokofyeva, M. Seon, J. Vanbuskirk, M. Holtz, S. A. Nikishin, N. N. Faleev, H. Temkin, and S. Zollner, Phys. Rev. B **63**, 125313 (2001).
[18] M. Kuball, Surf. Interface Anal. **31**, 987 (2001).
[19] C. A. Arguello, D. L. Rousseau, and S. P. d. S. Porto, Phys. Rev. **181**, 1351 (1969).
[20] C. Xiong, W. H. P. Pernice, X. Sun, C. Schuck, K. Y. Fong, and H. X. Tang, New J. Phys. **14**, 095014 (2012).
[21] W. H. P. Pernice, C. Xiong, and H. X. Tang, Opt. Express **20**, 12261 (2012).
[22] P. T. Lin, H. Jung, L. C. Kimerling, A. Agarwal, and H. X. Tang, Laser Photon. Rev. **8**, L23 (2014).
[23] W. H. P. Pernice, C. Xiong, C. Schuck, and H. X. Tang, Appl.Phys. Lett. **100**, 223501 (2012).
[24] C. Xiong, W. H. P. Pernice, and H. X. Tang, Nano Lett. **12**, 3562 (2012).
[25] H. Jung, C. Xiong, K. Y. Fong, X. F. Zhang, and H. X. Tang, Opt. Lett. **38**, 2810 (2013).
[26] X. W. Liu, C. Z. Sun, B. Xiong, J. Wang, L. Wang, Y. J. Han, Z. B. Hao, H. T. Li, Y. Luo, J. C. Yan, T. B. Wei, Y. Zhang, and J. X. Wang, Opt. Lett. **41**, 3599 (2016).
[27] J. C. Yan, J. X. Wang, P. P. Cong, L. L. Sun, N. X. Liu, Z. Liu, C. Zhao, and J. M. Li, Phys. Status Solidi C **8**, 461 (2011).
[28] K. Dovidenko, S. Oktyabrsky, J. Narayan, and M. Razeghi, J. Appl. Phys. **79**, 2439 (1996).
[29] X. W. Liu, C. Z. Sun, B. Xiong, L. Niu, Z. B. Hao, Y. J. Han, and Y. Luo, Vacuum **116**, 158 (2015).
[30] W.-L. Chen and Y.-C. Chang, J. Electron. Mater. **37**, 1064 (2008).
[31] R. W. Hellwarth, Phys. Rev. **130**, 1850 (1963).
[32] B. Jalali, V. Raghunathan, R. Shori, S. Fathpour, D. Dimitropoulos, and O. Stafsudd, IEEE J. Sel. Top. Quant. **12**, 1618 (2006).
[33] R. F. Begley, A. B. Harvey, and R. L. Byer, Appl. Phys. Lett. **25**, 387 (1974).
[34] H. M. Pask, P. Dekker, R. P. Mildren, D. J. Spence, and J. A. Piper, Prog. Quant. Electron. **32**, 121 (2008).
[35] N. Vermeulen, J. Lightwave Technol. **29**, 2180 (2011).
[36] J. M. Ralston and R. K. Chang, Phys. Rev. B **2**, 1858 (1970).
[37] T. Sander, S. Eisermann, B. K. Meyer, and P. J. Klar, Phys. Rev. B **85**, 165208 (2012).
[38] A. S. Liu, H. S. Rong, R. Jones, O. Cohen, D. Hak, and M. Paniccia, J. Lightwave Technol. **24**, 1440 (2006).
[39] R. Claps, D. Dimitropoulos, V. Raghunathan, Y. Han, and B. Jalali, Opt. Express **11**, 1731 (2003).

# Supplementary Material for "Continuous-wave Raman Lasing in Aluminum Nitride Microresonators"

Xianwen Liu,[1] Changzheng Sun,[1] Bing Xiong,[1] Lai Wang,[1] Jian Wang,[1] Yanjun Han,[1] Zhibiao Hao,[1] Hongtao Li,[1] Yi Luo,[1] Jianchang Yan,[2] Tongbo Wei,[2] Yun Zhang,[2] and Junxi Wang[2]

[1]*Tsinghua National Laboratory for Information Science and Technology, State Key Lab on Integrated Optoelectronics, Department of Electronic Engineering, University, Beijing 100084, China*

[2]*Research and Development Center for Semiconductor Lighting, Institute of Semiconductors, Chinese Academy of Sciences, Beijing 100083, China*

## I. CHARACTERIZATION OF THE DEVICE

To characterize the performance of the fabricated aluminum nitride (AlN) microring, transmission measurement is performed by employing a tunable laser (Santec TSL-510) as the light source and lensed fibers to couple light into and out of the chip. Meanwhile, a polarization controller is used to select polarized incident lights. The measurement results are shown in Figs. S1(a) and S1(b). High on-resonance extinction ratios and low off-resonance insertion losses of ~3.5 dB per facet are achieved for both transverse-magnetic (TM) and transverse-electric (TE) polarized incident lights, respectively.

To determine the quality factors ($Q$) of the microring, a high-resolution resonant linewidth measurement is implemented with the fine tuning function of the Santec TSL-510 tunable laser, which is frequency calibrated by beating with another tunable laser and detected with a high-speed photodetector in connection with an electrical spectrum analyzer. Furthermore, a Fano-like fit is employed to the measured data due to the existence of Fabry-Pérot interference induced by the chip end facet reflection [S1]. The measured linewidth and loaded $Q$ factors ($Q_L$) for fundamental $TM_0$ and $TE_0$ modes are illustrated in Figs. S1(c) and S1(d). According to the phase measurement, all these modes are under-coupled around 1550 nm. Thus, the intrinsic $Q$ factors ($Q_{int}$) are extracted to be ~2.3 and 1.5 million, corresponding to intracavity power enhancement factors of ~530 and 342, respectively.

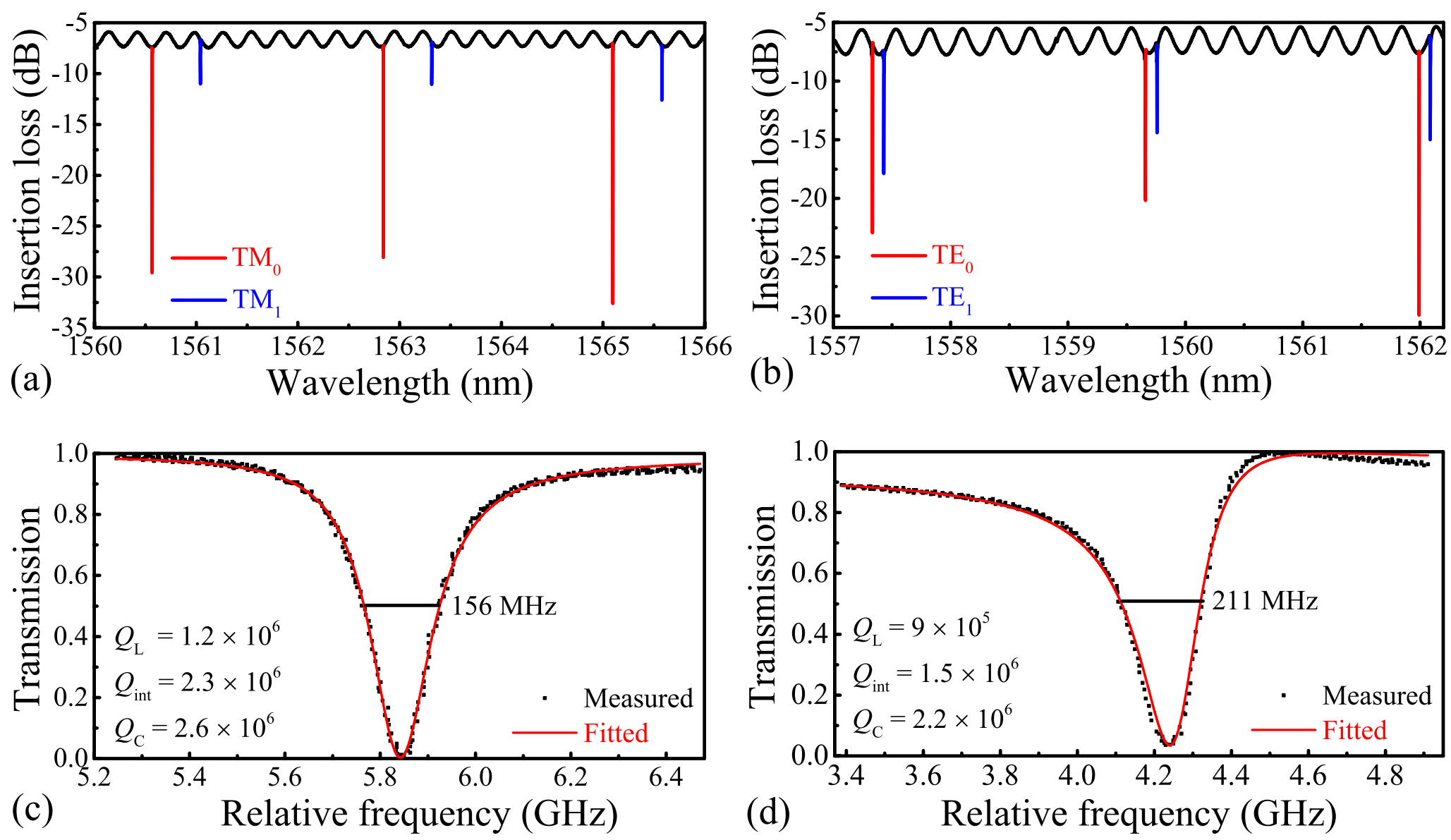

FIG. S1. (a) and (b) Transmission spectra for TM- and TE-polarized incident lights, respectively. The microring supports two TM and two TE modes, and the free spectrum range (FSR) for $TM_0$ and $TE_0$ modes are ~279 and 286 GHz, respectively. (c) and (d) High-resolution resonant linewidth measurements. A narrow full width at half maximum (FWHM) of ~156 and 211 MHz is obtained for $TM_0$ and $TE_0$ modes at ~1560.6 and 1559.7 nm, revealing loaded quality factors ($Q_L$) up to ~1.2 and 0.9 million, respectively.

## II. SPONTANEOUS RAMAN SPECTROSCOPY MEASUREMENT

Backscattering Raman spectroscopy measurement is implemented at room temperature with a Raman spectrometer (LabRAM HR Evolution, HORIBA). The specimen is the cleaved bus waveguide (for coupling light into the microring in the main text), exposing the $m$-plane $(10\bar{1}0)$ facet of AlN [Fig. S2]. A linearly-polarized (along $z$ axis) 532 nm laser is used as the excitation source, whereas $y$-polarized incident light is realized by inserting a halfwave plate into the incident beam path. The measurement range is 200−1000 cm$^{-1}$, and a high resolution of ~0.5 cm$^{-1}$ is ensured by adopting the 1800 gr/mm grating.

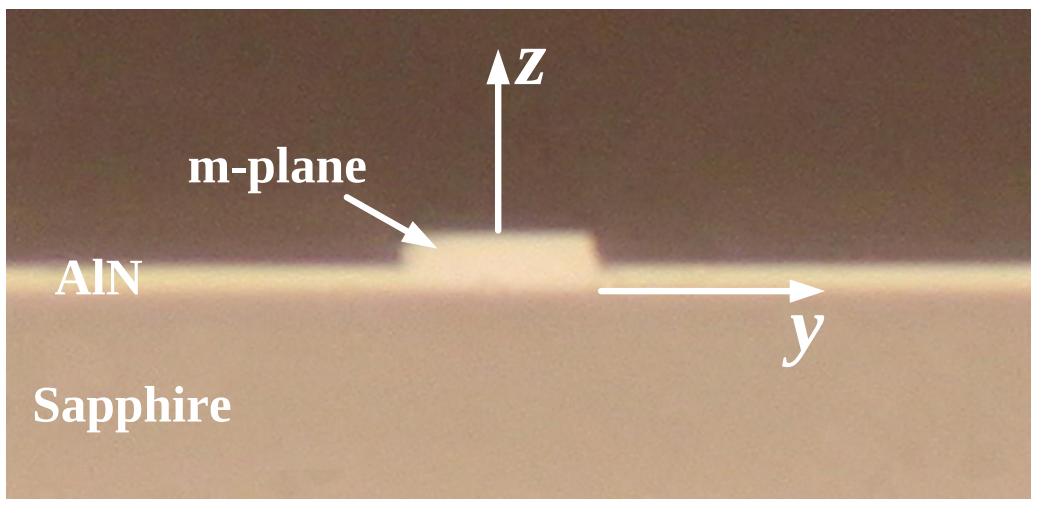

FIG. S2. Schematic illustration of the backscattering configuration at $m$-plane facet of AlN. The cleaved bus waveguide is used as the specimen shown in the above microphotograph, and an enhanced scattering efficiency is observed compared with that in bulk material. The Cartesian coordinate is established with the $z$ and $x$ axes perpendicular to the $c$- and $m$-plane of AlN, respectively.

## III. WIDEBAND DISCRETE TUNING OF THE STOKES LIGHT

Thanks to the low threshold (~8 mW) for TM-polarized pump, wideband discrete tuning of the 1$^{st}$ Stokes light is achieved by tuning the corresponding pump wavelength [Fig. S3]. Here, a C-band EDFA (1540−1570 nm) and a Santec TLS-510 tunable laser (1500−1630 nm) with high output power (~14.5 dBm) are employed as the pump sources, respectively. The absence of Raman lasing at certain longer wavelength is attributed to the limited pump power from the tunable laser as well as the degradation of $Q$ factor induced by mode splitting [S2].

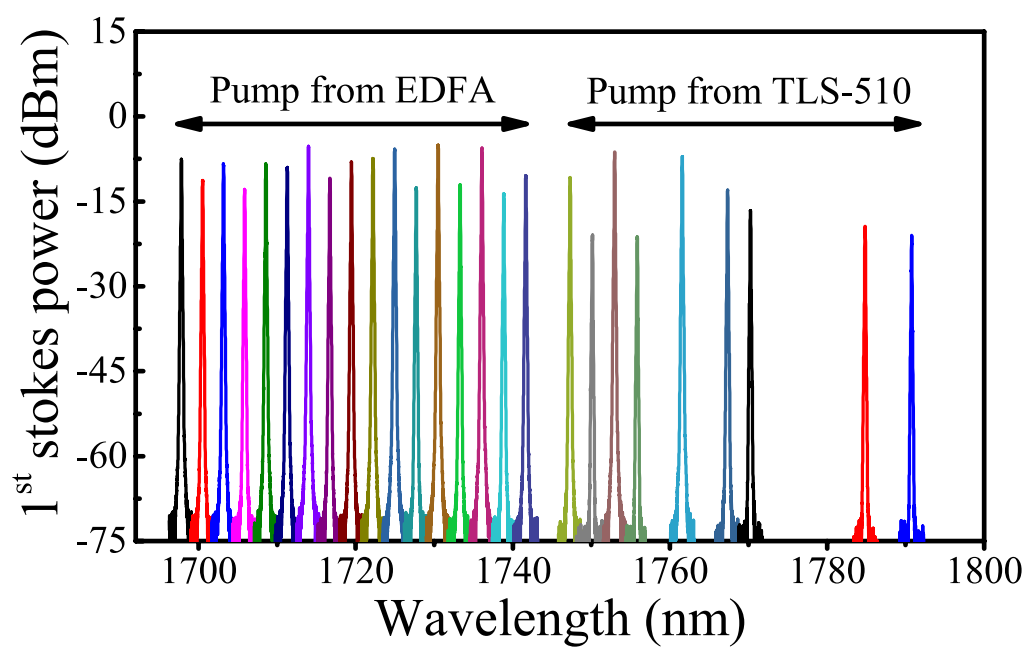

FIG. S3. Wideband discrete tuning of 1$^{st}$ Stokes light for TM-polarized pump. Here, 1$^{st}$ Stokes lasing from ~1700−1740 nm with single-FSR spacing is achieved with the pump from the EDFA. Meanwhile, Stokes lasing at longer wavelength (up to ~1790 nm) is demonstrated by utilizing the Santec TLS-510 tunable laser as the pump source.

## IV. SPONTANEOUS RAMAN SCATTERING EFFICIENCY ANALYSIS

As shown in the main manuscript, distinct Raman lasing threshold and slope efficiency are observed for TM and TE pump. Such lasing behavior can be related to the spontaneous Raman scattering efficiency $S$ of $A_1^{TO}$ and $E_2^{high}$ phonons. Typically, $S$ is given by the following expression [S3–S5]:

$$S = S_0 |\boldsymbol{e}_s \cdot \boldsymbol{R} \cdot \boldsymbol{e}_i|^2 \tag{S1}$$

where $S_0$ is a constant of proportionality, $\boldsymbol{R}$ represents the Raman tensor, $\boldsymbol{e}_i$ and $\boldsymbol{e}_s$ describe the polarization vectors of incident and scattered lights, respectively.

The complex Raman tensors for $A_1$, $E_1$, and $E_2$ phonons for wurtzite AlN are listed as follows [S3, S4]:

$$\boldsymbol{R}[\mathrm{A}_1] = \begin{pmatrix} |a|e^{i\psi_a} & 0 & 0 \\ 0 & |a|e^{i\psi_a} & 0 \\ 0 & 0 & |b|e^{i\psi_b} \end{pmatrix} \tag{S2}$$

$$\boldsymbol{R}[\mathrm{E}_1^x] = \begin{pmatrix} 0 & 0 & c \\ 0 & 0 & 0 \\ c & 0 & 0 \end{pmatrix}, \quad \boldsymbol{R}[\mathrm{E}_1^y] = \begin{pmatrix} 0 & 0 & 0 \\ 0 & 0 & c \\ 0 & c & 0 \end{pmatrix} \tag{S3}$$

$$\boldsymbol{R}[\mathrm{E}_2] = \begin{pmatrix} d & 0 & 0 \\ 0 & -d & 0 \\ 0 & 0 & 0 \end{pmatrix} + \begin{pmatrix} 0 & d & 0 \\ d & 0 & 0 \\ 0 & 0 & 0 \end{pmatrix} \tag{S4}$$

where $a$, $b$, $c$ and $d$ are the tensor matrix elements, while $\psi_a$ and $\psi_b$ describe the phases of $a$ and $b$, respectively. For the sake of simplicity, the tensors for $E_1$ and $E_2$ phonons are chosen to be real due to their equal nonzero tensor elements. Meanwhile, the Raman tensor for $E_2$ phonon is divided into two matrixes attributed to its twofold degeneracy.

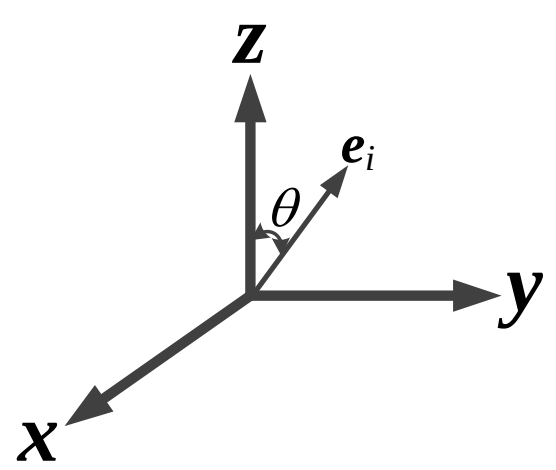

FIG. S4. Cartesian coordinates with $z$ and $x$ axes perpendicular to the $c$- and $m$-plane facets of AlN. The incident light is propagating along $x$ direction with a unit polarization vector of $\boldsymbol{e}_i$, which forms an angle of $\theta$ with respect to the $z$ axis.

First, we investigate the variation of $S$ for the incident light perpendicular to $m$-plane of AlN (similar to that in bus waveguide of Fig. S2). After defining $\boldsymbol{e}_i$ with a rotation angle $\theta$ against the $z$ axis [Fig. S4], $\boldsymbol{e}_i$ and $\boldsymbol{e}_s$ can thereby be derived as:

$$\boldsymbol{e}_i = \begin{pmatrix} 0 \\ \sin\theta \\ \cos\theta \end{pmatrix} \tag{S5}$$

$$\boldsymbol{e}_S^{\parallel} = \begin{pmatrix} 0 \\ \sin\theta \\ \cos\theta \end{pmatrix}, \quad \boldsymbol{e}_S^{\perp} = \begin{pmatrix} 0 \\ \cos\theta \\ -\sin\theta \end{pmatrix} \tag{S6}$$

where $\boldsymbol{e}_s^{\parallel}$ and $\boldsymbol{e}_s^{\perp}$ represent the polarization of scattered lights parallel and perpendicular to that of the incident light, respectively. For $\boldsymbol{e}_i \parallel \boldsymbol{e}_s$ case:

$$S^{\parallel}(\mathrm{A}_1) = S_0[|a|^2\sin^4\theta + |b|^2\cos^4\theta + \frac{1}{2}|a|\cdot|b|\sin^2(2\theta)\cos(\psi_a - \psi_b)] \tag{S7}$$

$$S^{\parallel}(\mathrm{E}_1) = S_0|c|^2\sin^2(2\theta) \tag{S8}$$

$$S^{\parallel}(\mathrm{E}_2) = S_0|d|^2\sin^4(\theta) \tag{S9}$$

whereas for $\boldsymbol{e}_i \perp \boldsymbol{e}_s$ case:

$$S^{\perp}(\mathrm{A}_1) = \frac{S_0}{4}[|a|^2 + |b|^2 - 2|a|\cdot|b|\cos(\psi_a - \psi_b)]\sin^2(2\theta) \tag{S10}$$

$$S^{\perp}(\mathrm{E}_1) = S_0|c|^2\cos^2(2\theta) \tag{S11}$$

$$S^{\perp}(\mathrm{E_2}) = \frac{S_0}{4}|d|^2 \sin^2(2\theta) \tag{S12}$$

For scattered lights with unpolarized detection, $S = S^{\parallel} + S^{\perp}$, which is analogous to the Raman spectrum measurement in Fig. 2 of the main text.

Hence, for $x(z, \cdot)\bar{x}$ geometry ($\theta = 0°$), we have $S(\mathrm{A_1}) = S_0|b|^2$, $S(\mathrm{E_1}) = S_0|c|^2$ and the ratio $|b|^2/|c|^2 = 11.4$ is derived from the measured scattering efficiency in Fig. 2(a). While for $x(y, \cdot)\bar{x}$ geometry ($\theta = 90°$), we have $S(\mathrm{A_1}) = S_0|a|^2$, $S(\mathrm{E_1}) = S_0|c|^2$, $S(\mathrm{E_2}) = S_0|d|^2$ . Based on the data in Fig. 2(b), it is found that $|a|^2/|d|^2 = 0.61$, $|c|^2/|d|^2 = 0.9$. As a consequence, $|b|^2/|d|^2 = 10.3$. In other words, $S(\mathrm{A_1})$ for $z$-polarized pump is ~10.3 times higher than $S(\mathrm{E_2})$ for $y$-polarized one. This is also in good agreement with the directly extracted value from Figs. 2(a) and 2(b) of the main text. Meanwhile, it is noted that maximum scattering efficiencies for $\mathrm{A_1^{TO}}$ and $\mathrm{E_2^{high}}$ phonons are attainable when excited with $z$- and $y$-polarized incident lights, respectively [Fig. S5].

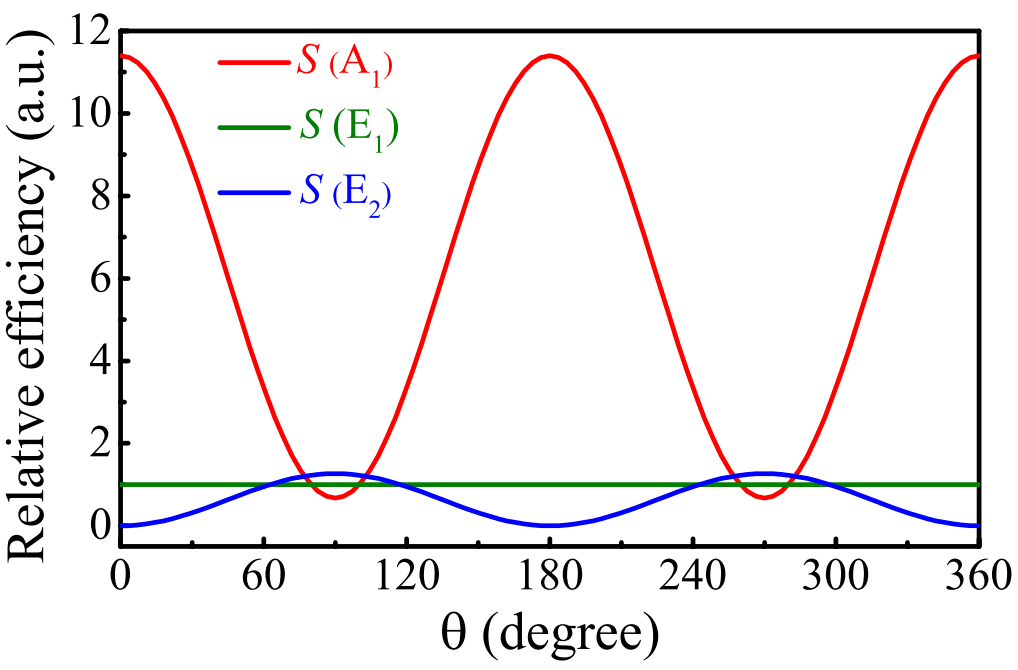

FIG. S5. Scattering efficienciy for $\mathrm{A_1}$, $\mathrm{E_1}$, $\mathrm{E_2}$ phonons versus the polarization angle $\theta$ of the incident light. Here, $S = S^{\parallel} + S^{\perp}$.

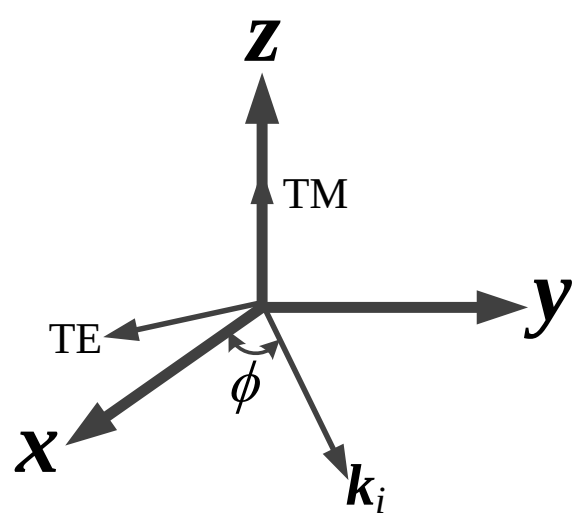

FIG. S6. Cartesian coordinates of arbitrary point within $c$-plane of AlN with $z$ and $x$ directions perpendicular to the $c$- and $m$-plane. $\boldsymbol{k}_i$ denotes the wave vector of the pump light, which has an angle of $\phi$ respect to $x$ axis. Here, TE and TM indicate the polarization directions.

To investigate the waveguide orientation dependence of Raman scattering efficiency, we consider the pump light propagating along arbitrary direction within the $c$-plane of AlN [Fig. S6]. By defining the wave vector of the pump light with an angle of $\phi$ against the $x$-axis, $\boldsymbol{e}_i$ and $\boldsymbol{e}_s$ can be derived as follows [S5]:
For TE-polarized case,

$$\boldsymbol{e}_{i,s} = \begin{pmatrix} \sin\phi \\ -\cos\phi \\ 0 \end{pmatrix} \tag{S13}$$

While for TM-polarized case,

$$\boldsymbol{e}_{i,s} = \begin{pmatrix} 0 \\ 0 \\ 1 \end{pmatrix} \tag{S14}$$

Thus, based on the Eq. S1, the scattering efficiencies within $c$-plane of AlN for TE- and TM-polarized lights are given as:

$$\mathrm{A_1\ phonon:}\quad \begin{cases} S^{\mathrm{TE-TE}}(\mathrm{A_1}) = S_0|a|^2 \\ S^{\mathrm{TM-TM}}(\mathrm{A_1}) = S_0|b|^2 \\ S^{\mathrm{TE-TM}}(\mathrm{A_1}) = S^{\mathrm{TM-TE}}(\mathrm{A_1}) = 0 \end{cases} \tag{S15}$$

$$\mathrm{E_1\ phonon:}\quad \begin{cases} S^{\mathrm{TE-TE}}(\mathrm{E_1}) = 0 \\ S^{\mathrm{TM-TM}}(\mathrm{E_1}) = 0 \\ S^{\mathrm{TE-TM}}(\mathrm{E_1}) = S^{\mathrm{TM-TE}}(\mathrm{E_1}) = S_0|c|^2 \end{cases} \tag{S16}$$

$$\mathrm{E_2\ phonon:}\quad \begin{cases} S^{\mathrm{TE-TE}}(\mathrm{E_2}) = S_0|d|^2 \\ S^{\mathrm{TM-TM}}(\mathrm{E_2}) = 0 \\ S^{\mathrm{TE-TM}}(\mathrm{E_2}) = S^{\mathrm{TM-TE}}(\mathrm{E_2}) = 0 \end{cases} \tag{S17}$$

where the superscript in $S$ describes the polarization of pump and Stokes lights, respectively. Based on the above calculation, it is concluded that for TM or TE pump light propagating within $c$-plane of AlN, the scattering efficiency $S$ of the optic phonons is independent of the waveguide orientation. Meanwhile, maximum $S$ for $\mathrm{A_1^{TO}}$ and $\mathrm{E_2^{high}}$ phonons are attainable for these two polarized pump [Fig. S5].

## V. THE ESTIMATION OF STIMULATED RAMAN GAIN COEFFICIENT

Based on the measured 1st Stokes lasing threshold $P_{\mathrm{th}}$ in the main manuscript, the Raman gain coefficient $g_R$ can be extracted according to the following expression [S6, S7]:

$$P_{\mathrm{th}} = \frac{\pi^2 n_P^2 \cdot V_{\mathrm{eff},P}}{\lambda_P \lambda_R \cdot g_R \cdot Q_{\mathrm{int},P} \cdot Q_{\mathrm{int},R}} \cdot \frac{(1+K_P)^2}{K_P} \cdot (1+K_R) \tag{S18}$$

where $n$, $\lambda$, $Q_{\mathrm{int}}$ describe the refractive index, wavelength, and intrinsic $Q$ factor, respectively. The subscript $P$ and $R$ denote the parameters at pump and Raman modes, respectively.

The effective mode volume $V_{\mathrm{eff},P}$ is calculated to be $\sim 1.12\times10^{-15}$ m$^3$ with finite element method (FEM), and $K$ is the coupling parameter determined by the ratio between $Q_{\mathrm{int}}$ and the coupling $Q$ factor ($Q_C$). The $K_P$ is obtained from the extracted $Q$ factors illustrated in Figs. S1(c) and S1(d). To determine the $K_R$, we assume equal intracavity propagation losses for the pump and 1st Stokes lights (i.e., an equivalent $Q_{\mathrm{int}}$). Meanwhile, the wavelength-dependent $Q_C$ is calculated from the coupling coefficient obtained with FIMMPROP simulation [S8] in Fig. S7. Taking into account the observed 1st Stokes lasing threshold of ∼8 and 34 mW, the corresponding $g_R$ in AlN is thereby extracted to be ∼0.45 and 0.25 cm/GW for TM and TE pump, respectively.

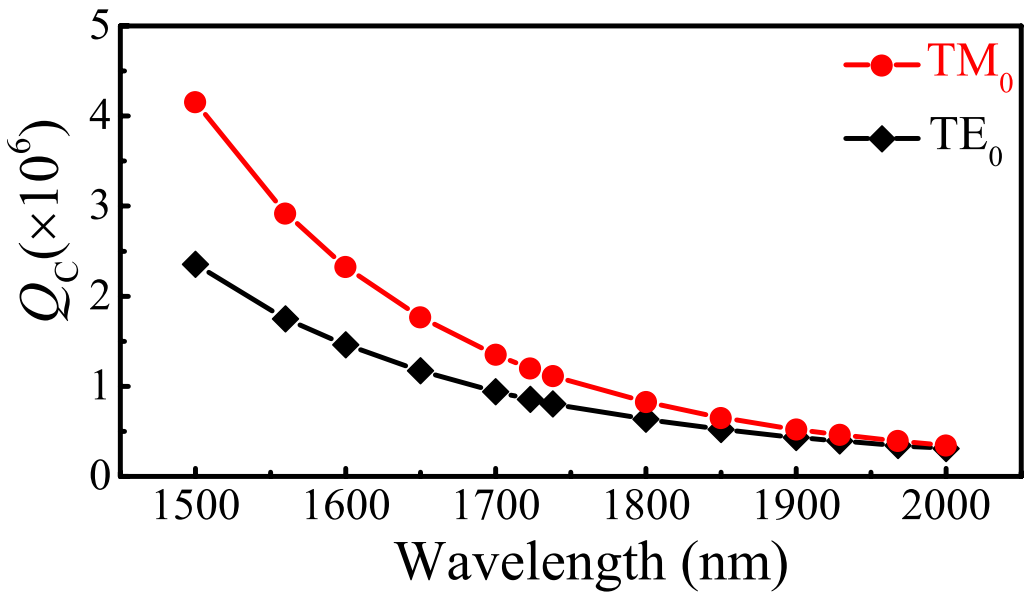

FIG. S7. Calculated $Q_C$ versus the wavelength for $\mathrm{TM_0}$ and $\mathrm{TE_0}$ modes, respectively. Note that the $Q_C$ decreases at the longer wavelength due to the increased coupling strength between the bus waveguide and microring. Additionally, it should be mentioned that the calculated $Q_C$ at ∼1560 nm is in accordance with the extracted values in Fig. S1, demonstrating its validity at 1st Stoke wavelength.

[S1] B. J. M. Hausmann, I. B. Bulu, P. B. Deotare, M. McCutcheon, V. Venkataraman, M. L. Markham, D. J. Twitchen, and M. Lončar, Nano Lett. **13**, 1898 (2013).

[S2] T. J. Kippenberg, S. M. Spillane, and K. J. Vahala, Opt. Lett. **27**, 1669 (2002).
[S3] L. E. McNeil, M. Grimsditch, and R. H. French, J. Am. Ceram. Soc. **76**, 1132 (1993).
[S4] T. Sander, S. Eisermann, B. K. Meyer, and P. J. Klar, Phys. Rev. B **85**, 165208 (2012).
[S5] A. S. Liu, H. S. Rong, R. Jones, O. Cohen, D. Hak, and M. Paniccia, J. Lightwave Technol. **24**, 1440 (2006).
[S6] S. M. Spillane, T. J. Kippenberg, and K. J. Vahala, Nature **415**, 621 (2002).
[S7] T. J. Kippenberg, S. M. Spillane, D. K. Armani, and K. J. Vahala, Opt. Lett. **29**, 1224 (2004).
[S8] X. W. Liu, C. Z. Sun, B. Xiong, J. Wang, L. Wang, Y. J. Han, Z. B. Hao, H. T. Li, Y. Luo, J. C. Yan, T. B. Wei, Y. Zhang, and J. X. Wang, Opt. Lett. **41**, 3599 (2016).